\documentclass[prx,amsmath,amssymb,aps,twocolumn]{revtex4-1}
\usepackage{graphicx}
\usepackage{bm}% bold math
\usepackage{natbib}
\usepackage[utf8]{inputenc}
\usepackage[squaren,Gray]{SIunits}
\usepackage{hyperref}
\usepackage{xcolor}

\renewcommand{\r}{\mathbf{r}}
\newcommand{\R}{\mathbf{R}}
\renewcommand{\k}{\mathbf{k}}

\begin{document}

\title{Bulk properties of honeycomb lattices of superconducting microwave resonators}

\author{Alexis Morvan}
\affiliation{Laboratoire de Physique des Solides, CNRS, Université Paris Saclay, Orsay, France}
\author{Mathieu Féchant}
\affiliation{Laboratoire de Physique des Solides, CNRS, Université Paris Saclay, Orsay, France}
\author{Gianluca Aiello}
\affiliation{Laboratoire de Physique des Solides, CNRS, Université Paris Saclay, Orsay, France}
\author{Julien Gabelli}
\affiliation{Laboratoire de Physique des Solides, CNRS, Université Paris Saclay, Orsay, France}
\author{Jérôme Estève}
\affiliation{Laboratoire de Physique des Solides, CNRS, Université Paris Saclay, Orsay, France}

\begin{abstract}
We have realized different honeycomb lattices for microwave photons in the 4 to \unit{8}{\giga\hertz} band using superconducting spiral resonators. Each lattice comprises a few hundred sites. Two designs have been studied, one leading to two bands touching at the Dirac points and one where a gap opens at the Dirac points. Using a scanning laser technique to image the eigenmodes of this new type of photonic lattices, we are able to reconstruct their band structure. The measured bands are in excellent agreement with \emph{ab initio} models that combine numerical simulations of the electromagnetic properties of the spiral resonator and analytical calculations.
\end{abstract}

\maketitle

Superconducting photonic lattices for microwave photons hold the promise to simulate the behaviour of strongly interacting bosons in tailored 1D and 2D lattices \cite{koch2009,koch2010,schmidt2013}. In such systems, the excitation are microwave photons with a typical frequency around $\unit{6}{GHz}$ stored in superconducting resonators with high quality factor. The main interest of this platform is the possibility to reach the strongly interacting regime using Josephson junctions as a non-linear element. But, Josephson junctions are prone to disorder, and the realization of large disorder free non-linear lattices remains a very difficult task. Nevertheless, many-body effects have already been demonstrated in small uni-dimensional lattices \cite{roushan2017, fitzpatrick2017, ma2018, guo2020stark, fedorov2020photon}.

Independently of controlling the interactions in the lattice, it seems also interesting to develop techniques allowing to probe lattices in the linear regime and to understand their properties \cite{underwood2016, kollar2019hyperbolic}. This research direction, towards the realization of engineered band structures for photons, catches up with the rapidly growing field of photonic lattices and topological photonics \cite{lu2014,khanikaev2017,ozawa2018}. In comparison to other photonic systems, the cryogenic environment and the low energy of the photons bring some technical difficulties to characterize the properties of such lattices. 

In this manuscript, we use a laser scanning imaging technique in order to map the spatial distribution of the resonant modes across the lattice. This technique is similar to the laser scanning microscopy developed to observe the current density in superconductor films \cite{culbertson1998} and has already been used to characterize single a superconducting resonator \cite{zhuravel2012,averkin2013} or coupled resonators \cite{wang2019mode}. It is an alternative to the scanning technique developed in \cite{underwood2016}. From a Fourier analysis of the spatial distribution of the modes, we reconstruct the dispersion relation of the lattice bands. We apply this method to three honeycomb lattices with different designs, where the $A$ and $B$ sites in the unit cell correspond to identical or different resonators. In one case, we observe two bands touching at the Dirac point, while in the latter, a gap opens at the Dirac points. These properties of the bands are similar to the ones for electrons in graphene, even though the two systems are not described by the same wave equation. In order to understand our lattices and pinpoint the difference with electronic systems, we have developed two approaches to predict the properties of the lattice. The first approach consists in projecting the Maxwell equations on the basis of the resonator modes, which corresponds to a coupled mode theory (CMT), a technique widely used in photonics \cite{huang1994}. The band structure of the lattice is then obtained in terms of overlap integrals between the electric and magnetic fields of neighbouring resonators. The second approach relies on more intensive numerical simulations to obtain an equivalent circuit to the lattice, from which the band structure can be calculated. These two \emph{ab initio} approaches allow us to reproduce the measured band structures with a very good accuracy. 

The manuscript is organized as follows. In the first part, we present the properties of the spiral resonator that is used as the site of the different lattices. In the second part, we present the measurement of the band structure of the three lattice designs. And the last part details the two models that we have developed to predict the band and mode structures of the lattices.

\section{Spiral resonator properties}\label{sec:spiral}
\begin{figure}
    \begin{center}
    \includegraphics[width=1.0\linewidth]{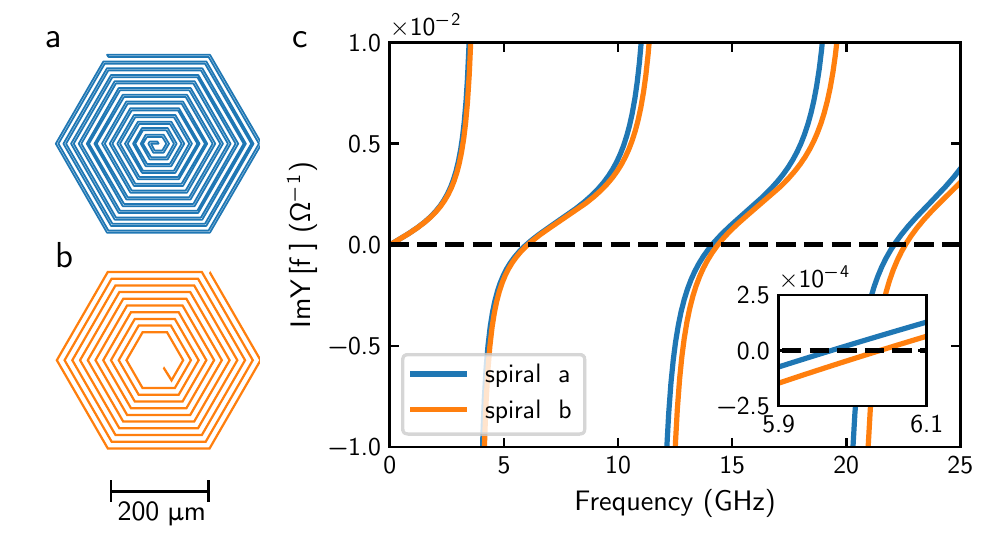}
    \caption{(a,b) Design of the two spirals used as the sites of our honeycomb lattices. The spiral shown in (a) consists of a \unit{8.6}{\milli\meter} long wire with a width of \unit{4.3}{\micro\meter} and a gap between adjacent turns of \unit{8.6}{\micro\meter}. The length of the (b) spiral is shorter resulting in a slightly higher resonance frequency. (c) Self-admittance of the two spirals calculated for a port located at the end of the wire at the center of the spiral. The resonances correspond to the zero-crossings. The inset shows a zoom close to the fundamental resonance. \label{fig:spiral_design_Y}}
    \end{center}
\end{figure}

The three different lattices studied in this manuscript are honeycomb lattices where each site consists of a superconducting spiral resonator as shown in figure~\ref{fig:spiral_design_Y}. The spiral is made of a \unit{4.3}{\micro\meter} wide Nb wire that is winded in an hexagonal pattern with an overall size of \unit{300}{\micro\meter}. The lattices are patterned through photo-lithography and reactive ion etching starting from a $\sim \unit{300}{\nano\meter}$ thick sputtered Nb layer on top of a silicon wafer. The design of the spiral was adjusted through numerical simulations such that the resonance frequency $\omega_0$ of the fundamental mode is close to \unit{2\pi \times 6}{\giga\hertz}. This frequency can be finely tuned by adjusting the length of the wire at the center of the spiral. For the experiments shown in this manuscript, we use the two designs shown in figure \ref{fig:spiral_design_Y}a,b that were chosen to obtain two sites with slightly different $\omega_0$ spaced by \unit{2\pi \times 120}{\mega\hertz}. The figure \ref{fig:spiral_design_Y}c shows the simulated self-admittance of the two designs, from which we deduce $\omega_0$ as well as the resonance frequencies of the higher order modes. The second mode is expected to resonate around \unit{14}{\giga\hertz} and plays no role in the experiments presented here, where all measurements are performed between 5 and \unit{7}{\giga\hertz}. 

\begin{figure}
    \begin{center}
    \includegraphics{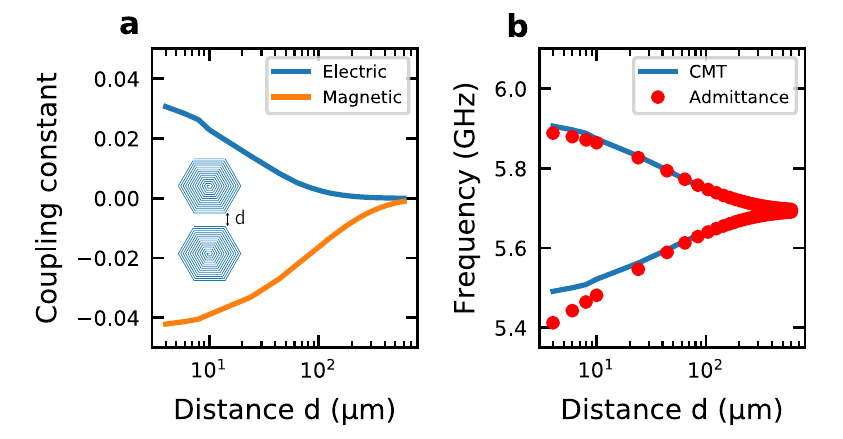}
    \caption{Coupling between two spiral resonators as shown in \ref{fig:spiral_design_Y}a a) Electric and magnetic coupling constants (see main text and Appendix A) as a function of $d$. b) Resonance frequencies $\omega_+$ and $\omega_-$ of the two coupled resonators as a function of $d$. Red dots correspond to the result of a numerical simulation that computes the admittance matrix of the two coupled resonators. The coupled mode theory prediction only uses the simulation of the electric and magnetic field created by a single resonator. It fails at short distances, where the resonator field is perturbed by the presence of the other resonator.   \label{fig:spiral_coupling}}
    \end{center}
\end{figure}

When two spirals are approached at a distance $d$ as shown in figure~\ref{fig:spiral_coupling}, the fundamental modes couple, giving rise to two resonances at frequencies $\omega_-$ and $\omega_+$. If the coupling is not too strong, the coupled modes can be expressed as a linear combination of the uncoupled modes. This approach that consists in using a reduced set of well chosen modes as a basis is known as the Coupled Mode Theory (CMT)\cite{powell2010, lomanets2012, elnaggar2015}. Here, we use a basic CMT with only one mode per resonator. The Maxwell equations projected in this basis lead to a linear system, whose eigenvalues correspond to the resonance frequencies $\omega_-$ and $\omega_+$. If the two resonators are identical with equal $\omega_0$, one obtains:
\begin{equation}
    \omega_\pm = \sqrt{\frac{1 \pm \kappa_e}{1 \pm \kappa_m}} \ \omega_0 \label{eq:omegapm}
\end{equation}
where $\kappa_e$ ($\kappa_m$) is the electric (magnetic) coupling constant. These coupling constants are proportional to the overlap of the electric (magnetic) fields of the two resonators. We define
\begin{align}
    D_{ij}  &= \int \epsilon(\r) \,  \mathbf{E}_i(\r) \cdot \mathbf{E}_j (\r) d^3 \r\\
    G_{ij}  &= \int  \mu_0 \,  \mathbf{H}_i(\r) \cdot \mathbf{H}_j (\r)  d^3 \r \label{eq:overlap}
\end{align}
where $E_i$ ($H_i$) is the electric (magnetic) field spatial dependence of the mode associated with the $i=1,2$ resonator. We assume these functions to be real. We explain in Appendix A how we compute these integrals from the charge and current distribution in the spiral that is obtained from the numerical simulation of a single resonator. In particular, one has to take into account charge and current images due to the dielectric interface at the surface of the sample and to the metallic ground plane below the sample. The coupling constants are then given by $\kappa_e = D_{12} / D_{11}$ and $ \kappa_m = G_{12}/G_{11}$. Figure~\ref{fig:spiral_coupling}a shows the evolution of $\kappa_e$ and $\kappa_m$ as a function of $d$ for two spirals shown in fig. 1a. Because $\kappa_e>0$ and $\kappa_m<0$, the magnetic and the electric couplings add up to increase the mode splitting, while the  mean frequency $(\omega_+-\omega_-)/2$ remains close to $\omega_0$. At large distances ($d>\unit{100}{\micro\meter}$), the magnetic coupling dominates, while at short distances both couplings are important. 

We compare the CMT predictions to a full numerical simulation of the two coupled resonators. From the simulation, we obtain the $2\times2$ admittance matrix $Y(\omega)$, which corresponds to the admittance matrix for two ports located at the ends of the two spirals (see figure). The two resonances $\omega_+$ and $\omega_-$ are then obtained as the zeros of $\det Y(\omega)$. The CMT predictions coincide with the ones of the full numerical simulation at large distances ($d>\unit{20}{\micro\meter}$) but fails at short distances as shown in  figure~\ref{fig:spiral_coupling}b. This is because, at short distances, the coupling is too strong and the coupled modes are not linear superposition of the uncoupled modes. In the following, we show results for lattices where $d=\unit{5}{\micro\meter}$ or $d=\unit{30}{\micro\meter}$. We therefore expect lattices with $d=\unit{30}{\micro\meter}$ to be well described by the CMT method, while lattices with $d=\unit{5}{\micro\meter}$ will require a more advanced model including the simulation of coupled sites. The half splitting $(\omega_+-\omega_-)/2$ obtained from the admittance matrix method is \unit{2\pi \times 200}{\mega\hertz} for $d=\unit{5}{\micro\meter}$ and \unit{2\pi \times 120}{\mega\hertz} for $d=\unit{30}{\micro\meter}$. This number gives an estimate of the nearest neighbour coupling amplitude in a tight-binding description of the lattice. 

\section{Lattice measurements}
\subsection{Density of states}
\begin{figure}
    \begin{center}
    \includegraphics[width=1.0\linewidth]{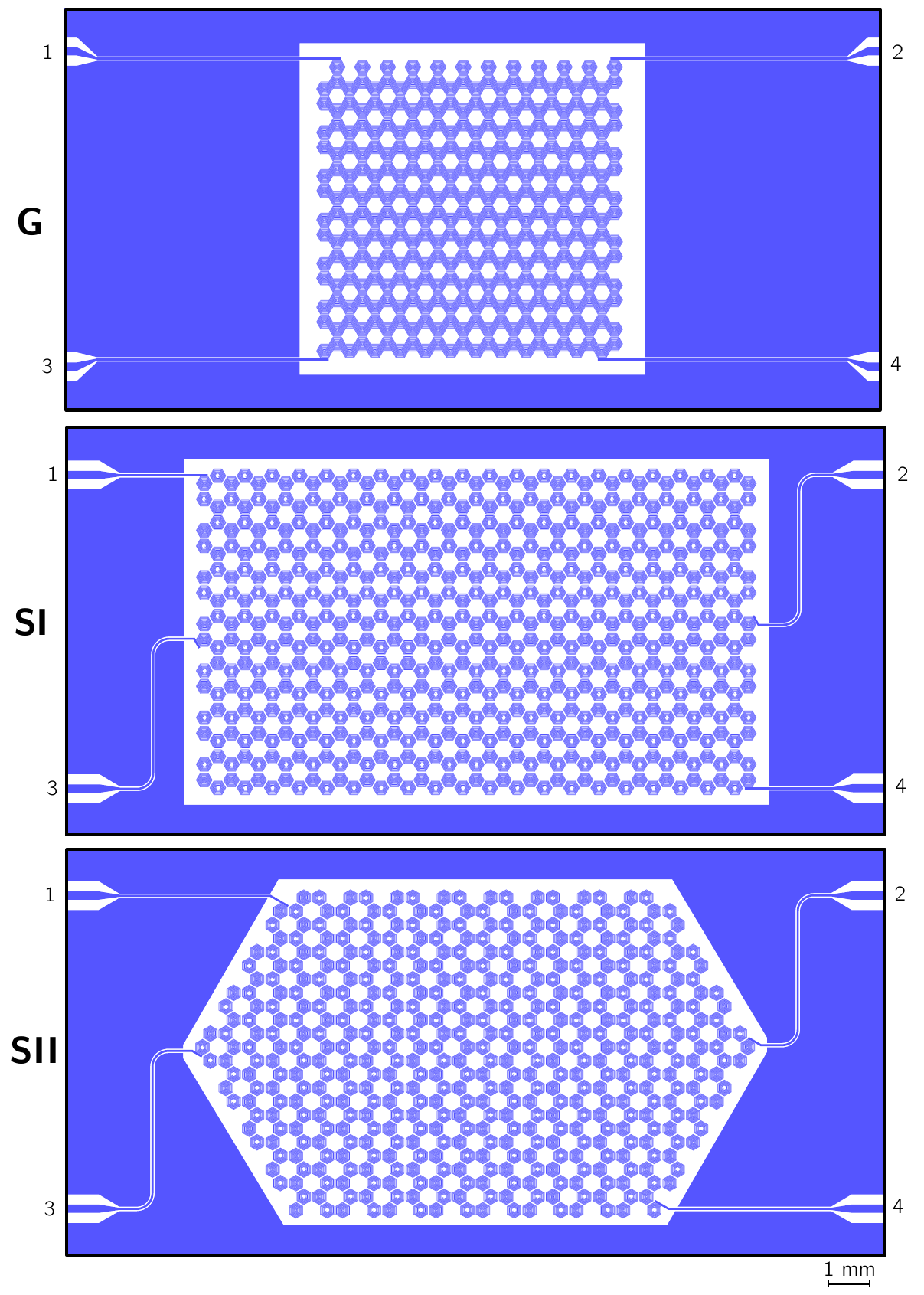}
    \caption{Design of the three different honeycomb lattices studied in this article. In the upper G design, all the sites are identical, while in the SI and SII designs, the A and B sites of the lattice correspond to two different resonators. Four coplanar waveguides are connected to single sites located on the edges of the lattice to probe the sample. \label{fig:designs}}
    \end{center}
\end{figure}

We have built honeycomb lattices consisting of a few hundred spiral resonators that fit on a $\unit{20\times 10}{\milli\meter\squared}$ sample, which is then mounted on the \unit{1}{\kelvin} stage of a dry dilution fridge. We have characterized three designs of honeycomb lattice labeled G,SI and SII, as shown in figure \ref{fig:designs}. The G design corresponds to the situation where all sites are identical, as in graphene, and are occupied by the spiral shown in figure 1a. The other two designs, SI and SII, correspond to the more general situation where the two nonequivalent A and B sites of the honeycomb lattice are occupied by the two different spirals shown in figure 1a and 1b. Because, to first approximation, the sites only differ by their onsite frequency, the SI and SII designs realize a so-called Semenoff insulator \cite{semenoff1984}. In the graphene case, the expected band structure consists of two bands touching at the two Dirac points, while in the Semenoff case a gap opens at the Dirac points. In the two Semenoff designs, a horizontal boundary divides the sample in two halves: The A (B) sites in the lower half are occupied by the resonators that occupy the B (A) sites in the upper half. The two halves correspond to the same infinite lattice and therefore have the same bulk properties. The role of this boundary is to create valley Hall boundary states as discussed in \cite{morvan2020}.
  
\begin{figure}
    \begin{center}
    \includegraphics[width=1.0\linewidth]{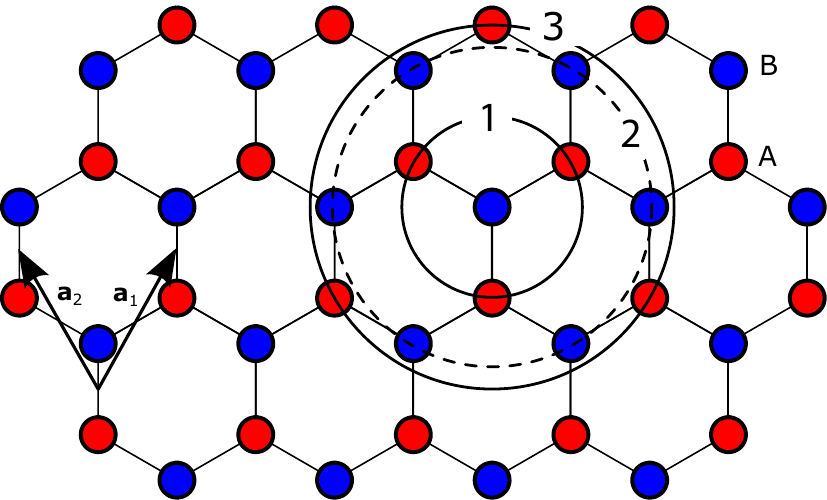}
    \caption{Honeycomb lattice with the two different sites A and B. The differents circles indicates the 1st, 2nd and 3rd neighbors sites. \label{fig:honeyxomb_lattice}}
    \end{center}
\end{figure}

\begin{figure}
    \begin{center}
    \includegraphics[width=1.0\linewidth]{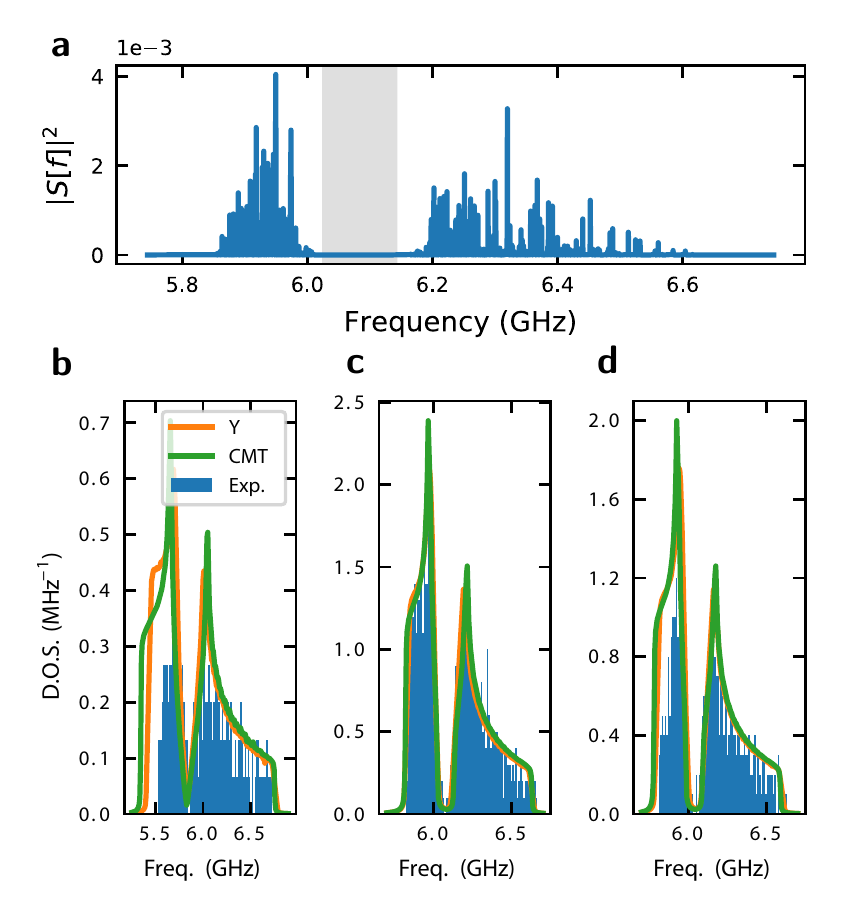}
    \caption{a) Transmission $|S_{xx}|^2$ for the SII sample. The lattice modes appear as sharp resonant peaks. The transmission on resonance is way below one indicating that the modes are under coupled: intrinsic loss dominate the coupling loss to the measurement waveguides. b,c,d) The blue data show the estimated DOS from transmission spectra similar as the one shown in a. %The orange curve is the calculated DOS of a tight binding model whose parameters are such that some remarkable points in the DOS (band edges and van Hove singularities) coincide with the measured ones. 
    The data are compared to the predictions of the two \emph{ab initio} models detailed in section \ref{sec:model}. The CMT model only uses the simulation of a single resonator, while the $Y$ matrix model relies on the simulation of the admittance matrix of a small lattice.
    The discrepancy between the data and the model is due to the fact that we miss some peaks and underestimate the DOS.
    \label{fig:DOS}}
    \end{center}
\end{figure}

 We probe the transmission through the lattice using four microwave ports that are connected to four coplanar waveguides. Each waveguide is coupled to a single site located on the edge of the lattice (see figure \ref{fig:designs}). We first characterize the bulk properties of the lattice by estimating the density of states (DOS) that we deduce from transmission measurements. Figure \ref{fig:DOS} shows a typical transmission spectrum obtained through a lattice of type SII. The lattice modes are visible as sharp resonant peaks. Depending on the sample, we identify between 43\% (G design) and 80\% (SI design) of the expected resonances. This number is limited by our signal to noise ratio and by the finite width of the peaks. Peaks that are too weakly coupled to the measurement ports are not identified as resonances and  peaks too close in frequency are counted as a single peak. By counting the number of peaks in a frequency window of 15\,MHz for the design G and 15\,MHz for the design SI and SII, we can however obtain an estimate of the DOS as shown in figure \ref{fig:DOS}. As expected, we observe two bands for all three samples, with a clear gap for the SI and SII samples. We can also identify the two van Hove singularities corresponding to the two maxima in the DOS of each band. %Following the method presented in \cite{bellec}, we use these remarkable points in the DOS to obtain the parameters of an effective tight-binding model (see Appendix B) that reproduces the experimental DOS as shown by the red lines in figure \ref{fig:DOS}. The model includes nearest-neighbour (NN), next nearest-neighbour (NNN) and next next nearest-neighbour (NNNN) couplings as well as an on-site frequency difference between the A and B sites for the S samples. The values are shown in the first row of table \ref{tab:coupling}.
The two \emph{ab initio} models to which we compare our data in the figure \ref{fig:DOS} only have a global offset frequency and are detailed in section \ref{sec:models}. The frequencies of the remarkable points observed in the DOS (band minima, maxima and Dirac point) are well reproduced, but the measured DOS is not in quantitative agreement with the model predictions because a significant fraction of the modes is missing from the measured DOS.
 
\subsection{Mode imaging and dispersion relation}
\begin{figure}
    \begin{center}
    \includegraphics[width=1.0\linewidth]{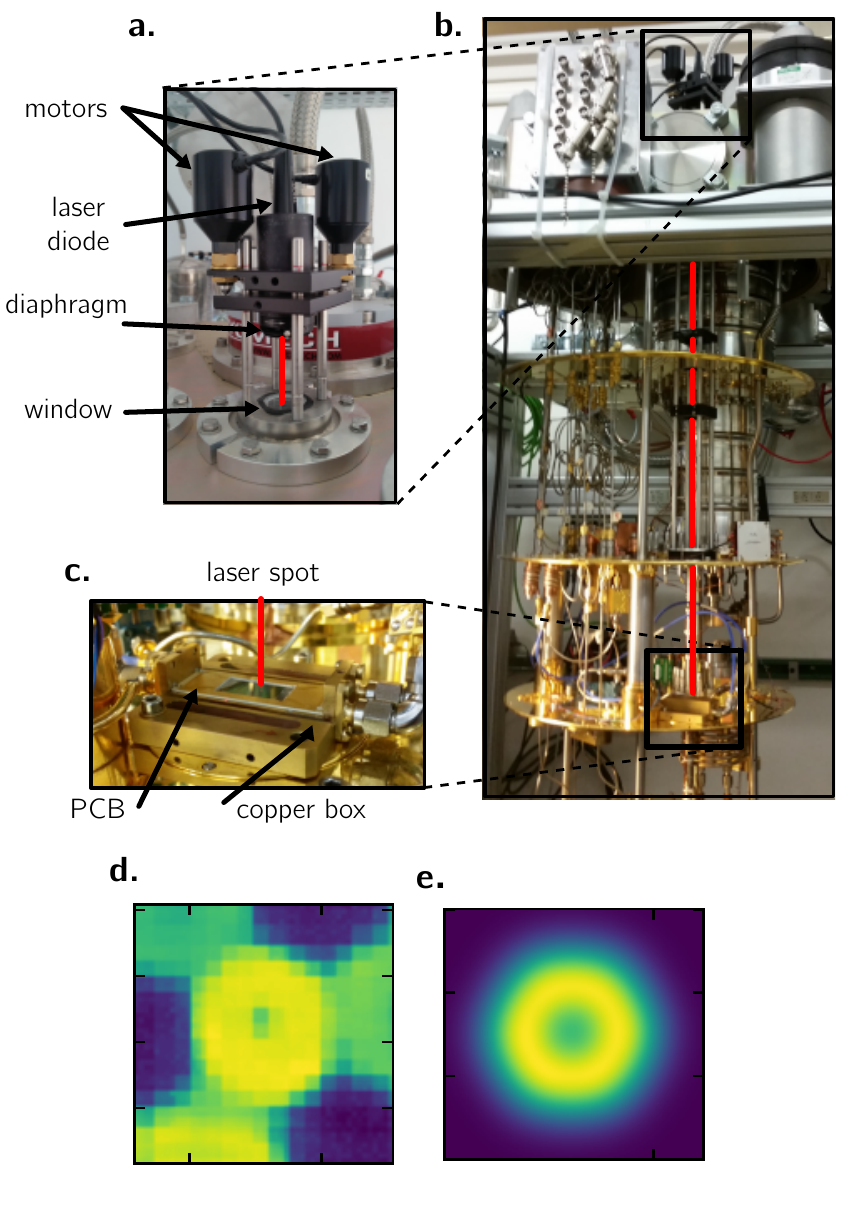}
    \caption{Mapping the spatial dependence of the lattice modes a,b,c) Cryogenic laser scanning setup. A collimated laser beam is mounted on a motorized mirror mount outside the dry refrigerator (a). The beam is relayed by two lenses to pass through openings in the \unit{50}{\kelvin} and the \unit{4}{\kelvin} shields (b). A final lens focuses the beam onto the sample that is clamped on the \unit{1}{\kelvin} stage (c). The tilt motion of the outer mirror mount results in a translation of the focused laser spot on the sample. d) Fine scan of a single resonator in the lattice. The image shows the transmission of one resonant peak as a function of the laser position. We compare this image to a convolution of the current intensity of the first mode of our spiral with a gaussian spot of waist $w$. Our waist size allows to discriminate each spiral and homogenise the response when the laser spot is on the resonator. \label{fig:setup}}
    \end{center}
\end{figure}

In order to further characterize the lattice, we use a laser scanning technique to map the spatial variation of the modes identified in the transmission spectra. This imaging technique allows us to obtain a partial information on the dispersion relation of the lattice modes. The measurement consists in monitoring the transmission of one (or many) mode while scanning a laser spot across the lattice. The experimental setup is shown in figure \ref{fig:setup}. This method has been previously used to map the spatial profile of the resonant modes of a single resonator \cite{zhuravel2012} or a chain of resonator. The optical setup was designed to obtain a laser waist on the sample of \unit{60}{\micro\meter}, which is much larger than the size of the spiral wire width and spacing but smaller than the overall resonator size. This allows us to average the mode distribution over each site, while keeping sufficient resolution to resolve adjacent sites. We observe that the main effect of the laser is to induce dissipation on the illuminated site as observed in \cite{zhuravel2006, zhuravel2012}. Because the modes are under-coupled to the probe ports, the laser induced increase of the loss results in a decrease of the mode transmission. To first order, the transmission drop for a given mode is proportional to the mode squared amplitude averaged over the illuminated area. In order to get rid of slow drifts and improve the signal to noise ratio, the laser intensity is modulated at a few \kilo\hertz. We then digitally demodulate the transmitted signal measured with a VNA and record the amplitude of the in phase signal as a function of the laser position.

Figure \ref{fig:setup}d shows a fine scan of one lattice site, which appears as a blurred hexagon with a central dip. This is consistent with the fact that the laser induced loss is maximal where the current density is large \cite{zhuravel2006} as shown by the simulation in figure 5d. We attribute a single value for the mode intensity per site by averaging over a few measurement points well inside the hexagon surrounding one site. We have checked that the final result is rather insensitive to the details of the averaging procedure. In order to optimize the acquisition time, different scanning techniques have been tested, continuous (as shown in figure 5e) or raster. We obtain best results with a raster scan consisting of six measurement points per site, while we monitor the transmission change of all the peaks in a frequency window of about \unit{1}{\giga\hertz}. This scan method allows us to image tens of modes in a single scan through the lattice.  

\onecolumngrid

\begin{figure}
    \begin{center}
    \includegraphics[width=1.0\linewidth]{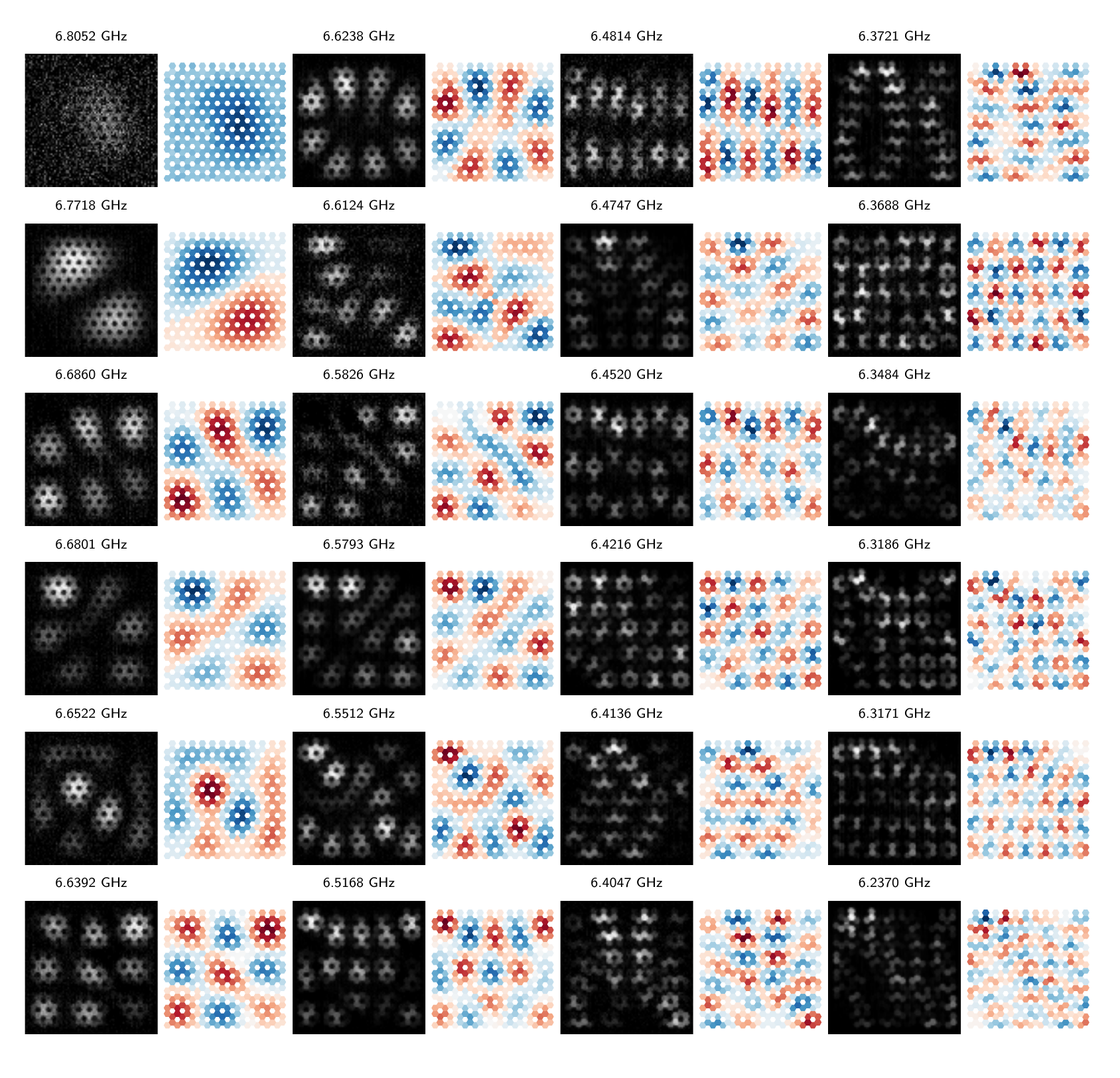}
    \caption{Measured mode spatial dependence for the G sample. The black and white images show the raw data coming from the measured transmission change at the frequency indicated above each image, which corresponds to a resonance peak identified in the transmission spectrum. This image is proportional to the mode intensity ac cross the lattice. The red and blue image shows the reconstructed signed mode amplitude (see main text and Appendix C). Dark blue (red) indicates a large negative (positive) amplitude.   
    \label{fig:modes_G}}
    \end{center}
\end{figure} 
\twocolumngrid

\onecolumngrid

\begin{figure}
    \begin{center}
    \includegraphics[width=1.0\linewidth]{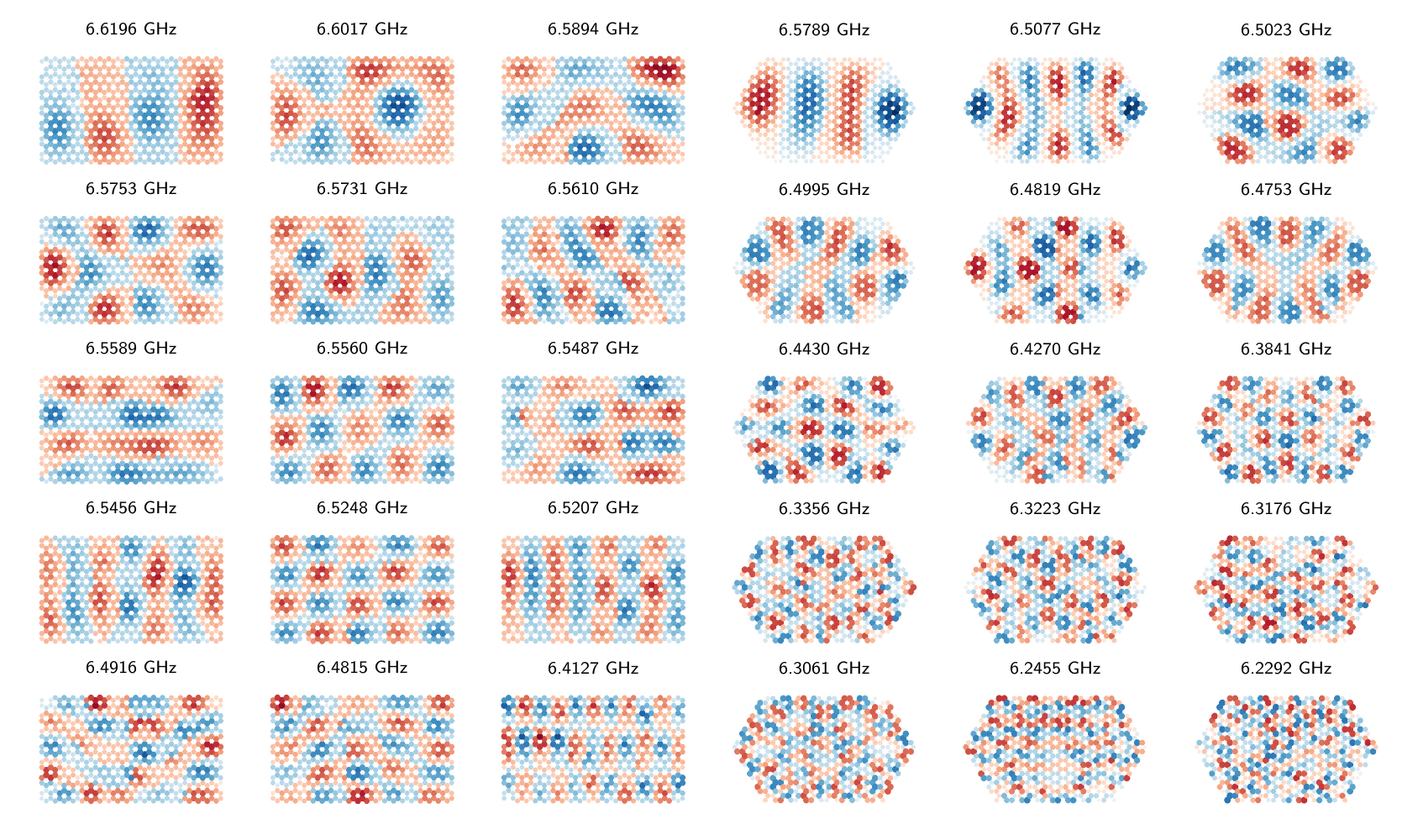}
    \caption{Measured mode spatial dependence for the SI and SII samples. Here, we only show the reconstructed signed mode amplitudes (see figure \ref{fig:modes_G}). The frequency of the mode is indicated above each image.
    \label{fig:modes_S}}
    \end{center}
\end{figure}
\twocolumngrid

Figures \ref{fig:modes_G} and \ref{fig:modes_S} show the results of the mode imaging technique applied to some modes of $G$, $SI$ and $SII$ samples. In order to attribute a wavevector to each measured mode, we have developed a reconstruction technique to deduce the signed mode amplitude from the measured intensity. We do so by supposing that the  mode is a linear combination of three known basis modes. As a basis, we choose the modes expected for a lattice having the same finite size geometry as the measured and described by a tight-binding model with only nearest neighbour coupling. We then look for the combination that matches best the measured mode intensity and use its sign to attribute a sign to the data. Details of the method are given in Appendix B. We then take the Fourier transform of the reconstructed signed mode amplitude. In the reciprocal space, several peaks lying on a circle are observed. This allows us to attribute a single value $k$ corresponding to the norm of a wavevector to each mode. Figure \ref{fig:BS} shows the dispersion relation $\omega(k)$ that we obtain through this analysis for the $G$ and $SI$ samples. The measured dispersion for the $SII$ sample is almost identical to the one of the $SI$ sample and is not shown here. We recover the two asymmetric bands observed in the DOS estimation. The mode dispersion is quadratic at small $|k|$ leading to an effective mass for the photons in the lattice which is on the order of () for the lower (upper) band. At larger $k$, the dispersion relation clearly deviates from a quadratic behaviour. For the G sample, we expect a linear dispersion around the Dirac points but we are not able to image a sufficiently large number of modes in this region to clearly reproduce this behaviour. This is due to the finite size of the sample and also to the radial averaging over the direction of the wavevector. For the SI sample, we observe that bands curve again in the opposite direction close to the Dirac points.
\begin{figure}
    \begin{center}
    \includegraphics[width=1.0\linewidth]{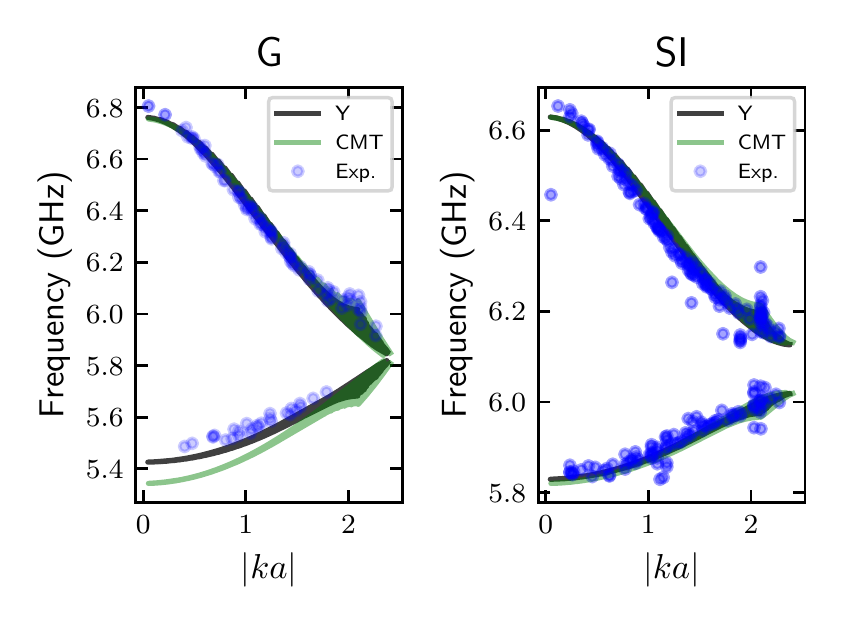}
    \caption{Band structure of the G and SI samples. For a given resonance frequency, the corresponding norm of the wavevector associated to the mode is obtained from the Fourier analysis of the measured spatial dependence as shown in figures \ref{fig:modes_G} and \ref{fig:modes_S}. %The data are compared to three models: a phenomenological tight-binding model in gray with parameters coming from the analysis of the measured DOS and two ab-initio models (see main text), one coming from a CMT analysis in green and the other from a $Y$ matrix calculation in black. 
    The measured dispersion relation is compared to the predictions of the two \emph{ab initio} models detailed in section \ref{sec:model} (see also figure \ref{fig:DOS}). The theoretical predictions are plotted as shaded areas, because, for a given $|k|$, the models predict different resonance frequencies depending on the orientation of the wavevector.
    \label{fig:BS}}
    \end{center}
\end{figure}

\section{Lattice models}\label{sec:model}
In comparison to the DOS, the measured dispersion shown in \label{fig:BS} is less affected by the fact that we are not able to probe all the modes. It allows us to precisely compare our data with two \emph{ab initio} models. Following the analysis of the coupling of two spiral resonators, we extend the CMT and the admittance matrix model to the case of an infinite lattice. The two models have no free parameters and differ in the following way: the CMT model solely relies on the simulation of the charge and current distribution of a single spiral at resonance, while the admittance matrix model simulates the admittance matrix of a cluster of a few coupled sites. As explained in \ref{sec:spiral}, the CMT model gives a clear physical picture of the coupling in the lattice in terms of overlap integrals between the different sites but fails when the coupling is too strong. As a consequence, the band structure of the S lattice is well predicted by the CMT but observe a discrepancy for the G lattice, which is more pronounced for the lower band. The $Y$ matrix calculation gives a more exact description of the lattice at any coupling but requires a more intensive numerical computation.  

\subsection{Coupled mode theory}
We follow the derivation of \cite{elnaggar2015} and adapt it to the specific case of coupled superconducting resonators. We only consider one mode per resonator and look for a solution to the Maxwell equations at a frequency $\omega$ as 
\begin{align}
    \mathbf{E}(\r,t) & = \sum_i a_i(t) \mathbf{E}_i(\r) \\
    \mathbf{H}(\r,t) & = \sum_i b_i(t) \mathbf{H}_i(\r)  
\end{align}
where $\mathbf{E}_i(\r)$ ($\mathbf{H}_i(\r)$) is the electric (magnetic) field of the mode associated to the resonator at site $i$. The two fields verify 
\begin{equation}
    \nabla \times \mathbf{E}_i = \mu_0 \omega_i H_i
\end{equation}
where $\omega_i$ is the resonance frequency of the mode of the resonator at site $i$. In addition to the overlap integrals defined in (\ref{eq:overlap}), we consider the integral
\begin{equation}
    M_{ij} = \int_S [ \mathbf{E}_i(\r) \times \mathbf{H}_j(\r) ] \cdot d\mathbf{S}
\end{equation}
where the surface $S$ corresponds to all the metallic boundaries in the circuit, which consists of the resonators and the walls of the box enclosing the sample. This integral can be rewritten in terms of the overlap integrals $D_{ij}$ and $G_{ij}$ assuming that 
\begin{equation}
    M_{ij} \approx \int_{S_i \cup S_j} [ \mathbf{E}_i(\r) \times \mathbf{H}_j(\r) ]\cdot d\mathbf{S} 
\end{equation}
where $S_i$ is the surface of the resonator at site $i$. The integral over $S_i$ is null because $\mathbf{E}_i(\r)$ is normal to the surface of the resonator. Using that the current density in the resonator at site $j$ is given by $\mathbf{J}_j(\r) = \mathbf{n} \times \mathbf{H}_j(\r)$, where $\mathbf{n}$ is the normal to the sample surface, we obtain
\begin{equation}
    M_{ij} \approx \int \mathbf{E}_i(\r) \cdot \mathbf{J}_j(\r) dV 
\end{equation}
Using this approximation and considering the volume integral of $\nabla \cdot (  \mathbf{E}_i \times \mathbf{H}_j)$, one obtains 
\begin{equation}
    2 M_{ij} = \omega_i G_{ij} - \omega_j D_{ij}  
\end{equation}
In matrix form, the last equation writes
\begin{equation}
    2M = \Omega G - D \Omega \label{eq:approxM}
\end{equation}
where $\Omega$ is the diagonal matrix with elements $\Omega_{ii}=\omega_i$. In the same way, the volume integrals of $\nabla \cdot (  \mathbf{H}_i \times \mathbf{E})$ and $\nabla \cdot (  \mathbf{E}_i \times \mathbf{H})$ lead to the following two equations 
\begin{align}
    G \dot{b} & = -\Omega D a \label{eq:Gb} \\
    2 M b & = \Omega G b - \omega D \dot{a} \label{eq:Mb}
\end{align}
Using the approximation (\ref{eq:approxM}) for $M$ and eliminating $b$, we look for periodic solutions at frequency $\omega$ and obtain the following eigenvalue problem
\begin{equation}
    \omega^2 a = \Omega G^{-1} \Omega D a  \label{eq:eigen}
\end{equation}
In the case of two resonators only, we recover the result given in (\ref{eq:omegapm}). The matrices $G$, $\Omega$ and $D$ are real symmetric matrices, but, in general, $\Omega G^{-1} \Omega D$ is not symmetric. But its eigenvalues are real and positive with non-orthogonal eigenvectors.

Equation (\ref{eq:eigen}) can be used to find the coupled mode frequencies of any ensemble of coupled resonators. In order to take advantage of the lattice periodicity, we relabel the $a_i$ and $b_i$ amplitudes in a given lattice cell as $a_\mu(\R)$ and $b_\mu(\R)$, where $\R$ is Bravais lattice vector identifying the position of the cell in the lattice and the $\mu$ index distinguishes the A and B sites. Using the same notation, we define the following overlap integrals between neighbouring sites in the lattice
\begin{align}
    \langle \mathbf{E}_\mu \mathbf{E}_\nu \rangle_\R &= \int \epsilon(\r') \, \mathbf{E}_\mu(r') \cdot \mathbf{E}_\nu (\r'-\R) \ d^3 \r'\\
    \langle \mathbf{H}_\mu \mathbf{H}_\nu \rangle_\R  &= \int \mu_0 \,  \mathbf{H}_\mu(\r') \cdot \mathbf{H}_\nu (\r'-\R) \  d^3 \r'
\end{align} 
We now look for periodic solutions over the lattice as
\begin{equation}
    a_\mu(R) = e^{i \k \cdot \R} a_\mu(\k) \ \ b_\mu(R)  =  e^{i \k \cdot \R} b_\mu(\k)
\end{equation}
and define the Fourier transform of the $D$ and $G$ overlap matrices 
\begin{align}
    D_{\mu \nu}(\k) & = \sum_\R e^{i \k \cdot \R} \langle \mathbf{E}_\mu  \mathbf{E}_\nu \rangle_\R \label{eq:Dk}\\
    G_{\mu \nu}(\k) & = \sum_\R e^{i \k \cdot \R} \langle \mathbf{H}_\mu  \mathbf{H}_\nu \rangle_\R \label{eq:Gk}
\end{align}
We suppose that the lattice contains $N$ cells and we look for solutions with periodic boundary conditions such that $\sum_\R e^{j \k \cdot \R} = N \delta_{\k,0}$, where $\k$ can take $N$ values in the first Brillouin zone. Equations (\ref{eq:Gb}) and (\ref{eq:Mb}) then lead to 
\begin{equation}
    \Omega G^{-1}(\k) \Omega D(\k) a(\k) = \omega^2(\k) a(\k) \label{eq:cmt}
\end{equation}
where $\Omega$ is now the diagonal matrix with the resonance frequencies of the A and B sites. The resulting eigenvalue problem (\ref{eq:cmt}) gives the dispersion of the two bands as a function of $\k$. 

We obtain the band structure of our lattices by including all couplings up to the third neighbour coupling (see figure \ref{fig:honeyxomb_lattice}) in equations (\ref{eq:Dk}) and (\ref{eq:Gk}) and solving (\ref{eq:cmt}). The results are shown in figure . In order to compare the obtained band structure with the one of a tight-binding model with the same range of coupling, we consider the case of the G lattice where the A and B sites have the same resonance frequency. In order to obtain a simple analytical formula, we neglect the dependence of the overlap integrals with the direction of separation (e.g. we suppose that $\langle \mathbf{E}_A \mathbf{E}_B \rangle_{\mathbf{a}_1}=\langle \mathbf{E}_A \mathbf{E}_B \rangle_{\mathbf{a}}$). With this approximation, a straightforward calculation shows that $D(\k)$ and $G(\k)$ are diagonal in the same basis, leading to the following dispersion relation $\omega_\pm(\mathbf{k})$
\begin{equation}
    \omega_\pm (\mathbf{k}) = \sqrt{\frac{1 + \kappa_e^{(2)} \, f_2(\k) \pm |\kappa_e^{(1)} \, f_1(\k)  + \kappa_e^{(3)} \, f_3(\k)|}{1 + \kappa_m^{(2)} \, f_2(\k) \pm |\kappa_m^{(1)} \, f_1(\k)  + \kappa_m^{(3)} \, f_3(\k)|}} \ \omega_0 \label{eq:omega_cmt}
\end{equation}
where the electric coupling constants are given by 
\begin{align*}
\kappa^{(1)}_e & = \langle \mathbf{E}_A \mathbf{E}_B \rangle_\mathbf{0} / \langle \mathbf{E}_A \mathbf{E}_A \rangle_\mathbf{0} \\
\kappa^{(2)}_e & =\langle \mathbf{E}_A \mathbf{E}_A \rangle_{\mathbf{a}_1} / \langle \mathbf{E}_A \mathbf{E}_A \rangle_\mathbf{0} \\
\kappa^{(3)}_e & =\langle \mathbf{E}_A \mathbf{E}_A \rangle_{\mathbf{a}_1+\mathbf{a}_2} / \langle \mathbf{E}_A \mathbf{E}_A \rangle_\mathbf{0}
\end{align*}
The magnetic couplings $\kappa_m^{(i)}$ are given by the same expressions in terms of magnetic field overlaps and the link functions $f_i(\k)$  correspond to 
\begin{align*}
    f_1(\mathbf{k}) &= 1 + e^{i \mathbf{k} \cdot \mathbf{a}_1} + e^{i \mathbf{k} \cdot \mathbf{a}_2} \\
    f_2(\mathbf{k}) &= 2 [ \, \cos \mathbf{k} \cdot \mathbf{a}_1 + \cos \mathbf{k} \cdot \mathbf{a}_2 + \cos \mathbf{k} \cdot (\mathbf{a}_1-\mathbf{a}_2) \, ]\\
    f_3(\mathbf{k}) &= e^{i \mathbf{k} \cdot (\mathbf{a}_1+\mathbf{a}_2)} + 2 \cos  \mathbf{k} \cdot (\mathbf{a}_1-\mathbf{a}_2) \\
\end{align*}
Expression (\ref{eq:omega_cmt}) can be compared to the tight-binding dispersion relation
\begin{equation}
    \omega^{\rm TB}_\pm (\mathbf{k}) = \omega_0 + t_2 \, f_2(\k) \pm |t_1 \, f_1(\k)  + t_3 \, f_3(\k)| \label{eq:omega_tb}
\end{equation}
which comes from the diagonalisation of the tight-binding Hamiltonian
\begin{equation}
    H^{\rm TB}(\mathbf{k}) = \begin{pmatrix}
    \omega_0 + t_2 f_2(\k) & t_1 f_1(\k) + t_3 f_3(\k) \\
    t_1 f_1^*(\k) + t_3 f_3^*(\k) & \omega_0 + t_2 f_2(\k)
    \end{pmatrix}
\end{equation}

The coupling $t_i$ corresponds to the $i$-th nearest neighbour coupling. The dispersion relation (\ref{eq:omega_cmt}) and (\ref{eq:omega_tb}) match if one keeps only nearest neighbour terms, performs a first order expansion in $\kappa_e^{(1)}$ and $\kappa_m^{(1)}$ and identifies $t_1$ to $\omega_0 (\kappa_{e,1}-\kappa_{m,1})/2$. Including 2nd and 3rd nearest neighbour overlap integrals lead to a dispersion relation that cannot be identified to its tight-binding counterpart with the same coupling range.

As mentioned in the introduction, a strong motivation to build lattices of superconducting resonators is to include non-linear elements in the lattice. For example, in the case considered here, one could design spirals with an empty area in the center where a transmon might be inserted and coupled to the spiral. In this context, it can be interesting to obtain an effective tight-binding Hamiltonian that describes the linear behaviour of the lattice. Using the CMT approach, we obtained the two equations of motion
\begin{align}
    \dot{a} & = \Omega b \\
    G \dot{b} & = -\Omega D a 
\end{align}
And the total energy in the system is given by
\begin{equation}
    H = \frac{1}{2} a^T D \, a + \frac{1}{2} b^T G \, b
\end{equation}
In order to identify $a$ and $b$ with the two quadratures of the resonator modes, we normalize the basis fields to 
\begin{equation}
    D_{ii} = G_{ii} = \hbar \omega_i
\end{equation}
With this definition, and in the absence of coupling between the resonators ($G=D=\hbar \Omega$), the equations of motion coincide with the Hamilton equations derived from $H$ assuming standard commutation rules between the mode quadratures, $[a_m,a_n]=0$, $[b_m,b_n]=0$ and $[a_m,b_n]=i \hbar \delta_{mn}$. In the following, we drop the $\hbar$ factors and set $\hbar=1$. 
When the overlap between adjacent sites are non-zero, the equations of motion do not coincide anymore with the naive Hamilton equations considering that the $a$'s and $b$'s commute. This comes from the fact that the basis used to project the Maxwell equations is not orthonormal, which modifies the commutation relations. This problem is well known in the calculation of electronic band structure using the LCAO method. In order to find an orthonormal basis, we adapt the Löwdin procedure used for electrons \cite{Aiken1980OnLO}. We first apply the canonical transformation $a \rightarrow \Omega^{1/2} a$ and $b \rightarrow \Omega^{-1/2} b$ to remove the $\Omega$ dependence in the equation of motion, which transforms the Hamiltonian to 
\begin{equation}
    H = \frac{1}{2} a^T \tilde{D} \, a + \frac{1}{2} b^T \tilde{G} \, b
\end{equation}
with $\tilde{D}=\Omega^{1/2}D\Omega^{1/2}$ and $\tilde{G}=\Omega^{-1/2}G\Omega^{-1/2}$. This step is not necessary if $\Omega$ is proportional to the identity. The equations of motion become $\dot{a}=b$ and $G\dot{b}=-Da$. We now apply the Löwdin transformation $a \rightarrow \tilde{G}^{-1/2} a$ and $b \rightarrow \tilde{G}^{-1/2} b$ in order to restore canonical commutation relations. The Hamiltonian becomes 
\begin{equation}
    H = \frac{1}{2} a^T \tilde{G}^{-1/2}  \tilde{D} \tilde{G}^{-1/2}\, a + \frac{1}{2} b^T \, b
\end{equation}
The equations of motion are modified to $\dot{a}=b$ and $\dot{b}=-\tilde{G}^{-1/2}  \tilde{D} \tilde{G}^{-1/2} a$, which now correspond to the Hamilton equations obtained with $H$ assuming canonical commutation relations. We finally apply the reverse canonical transformation $a \rightarrow \Omega^{-1/2} a$ and $b \rightarrow \Omega^{1/2} b$ to explicitly restore the $\Omega$ dependence. The final Hamiltonian is
\begin{equation}
    H = \frac{1}{2} a^T \Omega^{-1/2} \tilde{G}^{-1/2} \tilde{D} \tilde{G}^{-1/2} \Omega^{-1/2}\, a + \frac{1}{2} b^T \Omega \, b
\end{equation}
We can now introduce the bosonic operators $c = (a+ib)/\sqrt{2}$ and rewrite $H$ as 
\begin{equation}
    H =  \sum_i \frac{\omega_i}{2} (c_i^\dagger c_i + c_i c^\dagger_i)  + \frac{1}{2} (c + c^\dagger)^T J (c + c^\dagger)
\end{equation}
with $2J = \Omega^{-1/2} \tilde{G}^{-1/2} \tilde{D} \tilde{G}^{-1/2} \Omega^{-1/2} - \Omega$. A rotating wave approximation can then be performed to transform the last term to $c^T (J/2) c^\dagger + {\rm hc}$.

\subsection{Admittance matrix}
In order to model more accurately the G lattice, we extend the admittance matrix method introduced in section \ref{sec:spiral} to an infinite lattice of resonators and we also take into account the possibility to have $N \geq 1$ microwave ports per resonator. We introduce the voltage amplitude $V(\r')$ that is a complex vector with $2N$ components corresponding to the voltage at the $2N$ ports describing the voltage in the lattice cell at position $\r'$. At a given frequency $\omega$, the Kirchhoff's circuit laws describing the lattice can be written
\begin{equation}
	\sum_{\r}  Y_\r(\omega) V(\r'+\r) = 0
\end{equation} 
where the sum over $\mathbf{r}$ is over the points of the Bravais lattice. The matrix $Y_0(\omega)$ is the admittance matrix between the ports belonging to the same cell, while $Y_\r(\omega)$ is the admittance matrix between the ports corresponding to two cells separated by $\r$, it has the property $Y_{-\r}(\omega) = Y_{\r}^T(\omega)$. We look for a periodic solution $V(\r) = V(\k) e^{i \k \r}$ and obtain
\begin{equation}
	\sum_{\r} Y_\r(\omega) e^{i \k \r} V(\k) = 0
	\label{eq:KVL_kspace}
\end{equation}
where $V(\k)$ is a $2N$ element vector. The dispersion relation is then obtained by solving
\begin{equation}
    \det \left( \sum_r  Y_\r(\omega) e^{i \k \r} \right)=0
\end{equation}
The $Y_\r(\omega)$ matrix is obtained from the numerical simulation of a finite size lattice. The CMT calculations indicate that one should simulate a lattice with enough sites such that the central site is surrounded by all neighbouring sites up to the 3rd nearest neighbour. We therefore choose the geometry shown in figure with 16 sites. In order for the sum over $\r$ to converge rapidly when the distance between sites increases, the number of ports $N$ and their location on the spiral must be well chosen. We observe from the simulations that $N=1$ is not sufficient, while results obtained for $N>2$ and various port locations give the same results. The results of the final simulations shown in are obtained with $N=3$ ports located as shown in . 

Compared to the CMT model, the admittance matrix is numerically more demanding because it requires the simulation of a much larger (here 16) number of sites. The dispersion obtained for the S lattice confirms the prediction of the CMT model, which were already in good agreement with our experimental data. For the G lattice, we observe a deviation compared to the CMT model leading to a much better agreement with the measured dispersion. We can thus conclude that our lattice are well under control and that \emph{ab initio} electromagnetic simulation accurately describe their properties.

\vspace{5pt}

\begin{acknowledgments}
    The authors gratefully acknowledge the conversations and insights of Marco Aprili.
\end{acknowledgments}

\bibliography{main}%

\end{document}